\documentclass[twocolumn,apl,amsmath,amssymb,showpacs,superscriptaddress]{revtex4-2}
\usepackage{epsf}
\usepackage[utf8]{inputenc}
\usepackage{amsmath}
\usepackage{amsfonts}
\usepackage{amssymb}
\usepackage{makeidx}
\usepackage{color}
\usepackage{gensymb}
\usepackage{graphicx}
\usepackage{lipsum}
\usepackage{mathtools}
\usepackage{float}
 \usepackage{multirow}
\usepackage[hidelinks,colorlinks=true,linkcolor=blue,citecolor=blue]{hyperref}
\DeclareUnicodeCharacter{2212}{\textendash}

\begin{document}
\title{Structural properties, dielectric relaxation and impedance spectroscopy of NASICON type Na$_{3+x}$Zr$_{2-x}$Pr$_{x}$Si$_2$PO$_{\rm 12}$ ceramics}

\author{Ramcharan Meena}
\affiliation{Department of Physics, Indian Institute of Technology Delhi, Hauz Khas, New Delhi-110016, India}
\affiliation{Material Science Division, Inter-University Accelerator Center, Aruna Asaf Ali Road, New Delhi-110067, India}
\author{Rajendra S. Dhaka}
\email{rsdhaka@physics.iitd.ac.in}
\affiliation{Department of Physics, Indian Institute of Technology Delhi, Hauz Khas, New Delhi-110016, India}

\date{\today}      

\begin{abstract}
We investigate the dielectric and impedance spectroscopic investigation of Pr-doped NASICON type Na$_{3+x}$Zr$_{2-x}$Pr$_{x}$Si$_2$PO$_{\rm 12}$ ($x=$ 0.05--0.2) samples as a function of temperature and frequency. The Rietveld refinement of x-ray diffraction patterns confirms the monoclinic phase having C2/c space groups for all the samples. The scanning electron microscopy shows the granular-like structure and energy dispersive x-ray analysis confirms the desired compositions. The temperature (90--400~K) and frequency (20 Hz-2 MHz) dependence of electric permittivity are explained using Maxwell-Wagner-Sillars (MWS) polarization and space charge polarization mechanisms. The dielectric relaxation shows nearly equal activation energy for all the samples with a non-Debye type of relaxation in the measured temperature range. The complex impedance analysis shows the presence of broad grain and grain boundary relaxation peaks. The stretched exponent analysis of electric modulus using the Kohlrausch-Williams-Watts (KWW) function further confirms the non-Debye type of relaxation. Moreover, scaling analysis of the electric modulus shows a similar type of relaxation for all the samples. The {\it a.c.} conductivity data are fitted using modified power law, where the temperature dependence of exponent ($s$) confirms the correlated barrier hopping (CBH) type conduction for all the samples. Our results indicate that the Pr doped NASICON samples are potential candidates for charge storage devices due to their large electric permittivity. 
 
\end{abstract}

\maketitle
\section{\noindent ~Introduction}

In recent years, the Na-based energy storage devices have attracted much attention due to their abundance and low cost as compared to Li-based systems \cite{Sapra_Wiley_2021, Chandra_EA_2020, Pati_JMCA_2022, Pati_JPS_24}. Among the large range, the NASICON (Sodium SuperIonic CONductor)-based materials having chemical formula Na$_{1+x}$Zr$_2$Si$_x$P$_{3-x}$O$_{\rm12}$ (0$\le$$x$$\le$3) are considered potential candidates for solid electrolyte in Na-ion batteries \cite{Dubey_AEM_21}. This is mainly due to their  excellent ionic conductivity ($\approx$10$^{-3}$ Scm$^{-1}$), high chemical and thermal stability with Na, three dimensional (3D) diffusion path for Na ion conduction and high stability against humid air \cite{Goodenough_MRB_1976, Hong_MRB_1976, Rao_SSI_21}. The other technological applications of NASICON materials are gas sensors, supercapacitors, ion-selective electrodes, and solid oxide fuel cells \cite{Rao_SSI_21, Fergus_JPS_06}. The lower density of NASICON materials makes it a good choice for weight reduction \cite{Rao_SSI_21, Singh_JES_21}. The crystal structure of NASICON materials depends on the chemical composition and has a monoclinic structure (C 2/c symmetry) for (1.8$\le$$x$$\le$2.2) otherwise, stabilized in a rhombohedral structure (R-3c symmetry) for all other values of $x$$\le$3. The monoclinic phase also transforms into the rhombohedral phase at higher temperatures around 430~K due to the shear deformation of unit cell \cite{Jolley_JACS_15, Rao_SSI_21, Singh_JES_21}. In Na$_{1+x}$Zr$_2$Si$_x$P$_{3-x}$O$_{\rm12}$, the lattice structure unit cell contains ZrO$_6$ octahedra, Si/PO$_4$ tetrahedra having four formula units per unit cell where $x=$ 2 exhibits the highest conductivity at room temperature in monoclinic phase \cite{Oh_AMI_19, Singh_Ionics_22}. In both structures (monoclinic and rhombohedral), Na-ions travel through the bottleneck triangular area formed by oxygen atoms of neighboring octahedral and tetrahedral units. The conductivity of NASICON materials is enhanced by increasing the bottleneck triangle area and providing a 3D transport path for Na-ion migration/ hopping. These (monoclinic and rhombohedral) structures contain four Na ions per unit cell having one Na$_1$ and three Na$_2$ atomic sites in the rhombohedral phase, where the Na$_2$ sites are further split into one Na$_2$ and two Na$_3$ (three-fold) sites in monoclinic phase. Out of these total available sites, 2/3 fraction are occupied and remaining 1/3 part is vacant. These vacant sites provide conductivity due to diffusion between Na$_1$--Na$_2$ and Na$_1$--Na$_3$ sites. The Na$_1$--Na$_2$ distance is optimized for a faster hopping rate by optimizing or tunning the lattice size. The conductivity of the NASCION materials can be altered by tunning the microstructure, crystalline size, and/or aliovalent doping to change the Si/P ratio \cite{Jolley_JACS_15, Rao_SSI_21, Singh_JES_21}. The distance between the neighboring available sites has to be optimized for faster hopping by tuning the lattice size. The Na-ions travel through the puckered hexagonal rings having alternative octahedral and tetrahedral edges. The {\it ab-inito} molecular dynamics show the correlated hopping dynamics followed by the Na ions, which can be  enhanced by increasing the Na concentration. The lower activation energy in the rhombohedral phase is due to more symmetric structure as compared to the distorted monoclinic phase \cite{Rao_SSI_21, Singh_Ionics_22}. 

Moreover, doping of divalent or trivalent atoms at the Zr site not only increases the concentration of Na mobile ions (as additional Na ions are needed to compensate the charge imbalance), but also modifies the triangular bottleneck area (decided by the ionic radius of the dopant element and its bond length) for Na ion conduction \cite{Rao_SSI_21, Wang_ESM_23}. The synthesis method and sintering temperature are two important factors in determining the bulk density and microstructure of the sintered samples. The sintering temperature should be carefully selected as above 1200\degree C lead to the formation of an insulating secondary phase of monoclinic zirconia (ZrO$_2$) at the grain boundary due to the volatilization of Na and P ions. The required sintering temperature can be lowered in different methods like rapid microwave sintering, ultrafast sintering, and doping of multivalent atoms \cite{Oh_AMI_19, Jing_CEJ_23, Wang_JPS_21, Wang_ESM_23}. It is crucial to investigate the dielectric properties of different doped NASICON ceramics for their possible use in charge storage devices. Though there are few high-temperature microwave absorption studies reported in literature \cite{Chen_ML_2018, Chen_JECS_18}, there are very few studies on the temperature dependence of dielectric relaxation and impedance spectroscopy of doped Na$_3$Zr$_{2}$Si$_2$PO$_{\rm 12}$ ceramics \cite{Dubey_AEM_21, Meena_CI_22}. More importantly, to the best of our knowledge there are no studies available in the literature on the temperature and frequency-dependent dielectric properties of trivalent-doped NASICON ceramics. 

Therefore, in this paper, we report the Structural properties, dielectric relaxation and impedance spectroscopic studies of Pr-doped NASION type Na$_{3+x}$Zr$_{2-x}$Pr$_{x}$Si$_2$PO$_{\rm 12}$ ($x=$ 0.05--0.2) ceramics. We find the grain and grain boundary relaxation peaks using the complex impedance analysis. The stretched exponent analysis of electric modulus using the Kohlrausch-Williams-Watts (KWW) function show the non-Debye type of relaxation. The analysis of {\it a.c.} conductivity data using modified power law and the temperature dependence of exponent ($s$) suggest the correlated barrier hopping (CBH) type conduction for all the samples.

\section{\noindent ~Experimental Details}

Polycrystalline Pr-doped samples with chemical composition of Na$_{3+x}$Zr$_{2-x}$Pr$_{x}$Si$_2$PO$_{\rm 12}$ ($x=$ 0.05--0.2) were synthesized using the solid-state reaction method. Firstly, a stoichiometric amount of Na$_2$CO$_3$ (purity 99.5\%), ZrO$_2$ (purity 99.5\%), Pr$_6$O$_{11}$ (purity 99.99\%), SiO$_2$ (purity 99\%) and NH$_4$H$_2$PO$_4$ (purity 99.9\%) were mixed using the agate mortar-postal. The Na$_2$CO$_3$ and NH$_4$H$_2$PO$_4$ powders were taken in 15\% excessive amounts to compensate for the loss of Na and P ions at high temperatures. The SiO$_2$ crystals were crushed into the powder form and preheated at 175$^\circ$C for 16 hrs to remove any absorbed moisture. All the chemical powders were mixed and grounded thoroughly for uniformity. The resultant mixed powder (grey color) was calcinated at 1100$^\circ$C for 12 hrs in the air. The calcinated powder, which became sky blue, was obtained in colloidal form. The calcinated powder was further grounded to have a uniform distribution of particles throughout the volume and pressed into the circular pellets of diameter 10~mm with a thickness of 1.2~mm using the hydraulic pressure (1000 PSI). These pellets were finally sintered at 1150$^\circ$C for 12 hrs with heating and cooling rates of $5^\circ$C/min.   
 
X-ray diffraction (XRD) having Bragg-Brentano geometry in the 2$\theta$ range of 10$^\circ$-80$^\circ$ with a step size of 0.0167$^\circ$ was used to determine the crystal structure and related phase of prepared samples. The patterns are recorded using PANalytical X’Pert-PRO having Cu-K$_\alpha$ radiation ($\lambda$=1.54~$A^\circ$) with working voltage and current of 40 kV and 30 mA, respectively. The Rietveld refinement of the collected XRD patterns was performed using the FullProf suite software \cite{Carvajal_CRNS_00}. The microscopic analysis was performed using the MIRA II LMH field emission scanning electron microscope (FE-SEM). The elemental composition is confirmed by the NCA PentaFET3 energy dispersive x-ray detector from Oxford attached with SEM. All frequency-dependent measurements were performed using an Agilent LCR meter (Model-E4980A) in parallel capacitance mode by the standard four-probe method. To avoid any error due to cable wiring capacitance and conductance open, short and cable length corrections (calibrations) were performed before starting the frequency-dependent measurements. The electrical connections on ceramic pellets were made using silver paint and dried at 200$^\circ$C for 2 hrs. These prepared pellets of Na$_{3+x}$Zr$_{2-x}$Pr$_{x}$Si$_2$PO$_{\rm 12}$ ($x=$ 0.05--0.2) acted as a dielectric medium and coated silver as metal electrodes in a capacitor configuration. The edge effect was minimized by keeping the fabricated electrode edge inside the sample boundary. The {\it parallel capacitance} ($C_P$), {\it dielectric loss} (D=tan$\delta$), {\it total impedance} ($\lvert Z \rvert$) and {\it phase angle} ($\theta$) measurements are performed in the frequency range of 20~Hz--2~MHz by keeping 1~V {\it a.c.} signal as an input perturbation. The temperature is varied from 90~K to 400~K using the Lakeshore temperature controller (Model-340) keeping the temperature stability of 100~mK with a stability time of 2 minutes having a fixed heating rate of  $2^\circ$C. All the frequency-dependent measurements were performed in the vacuum around $10^{-3}$ mbar. All samples were dipped in the liquid nitrogen dewar during the experiments for uniform cooling across the sample.     

\section{\noindent ~Results and discussion}

\begin{figure*}
    \centering
    \includegraphics[width=1.0\textwidth]{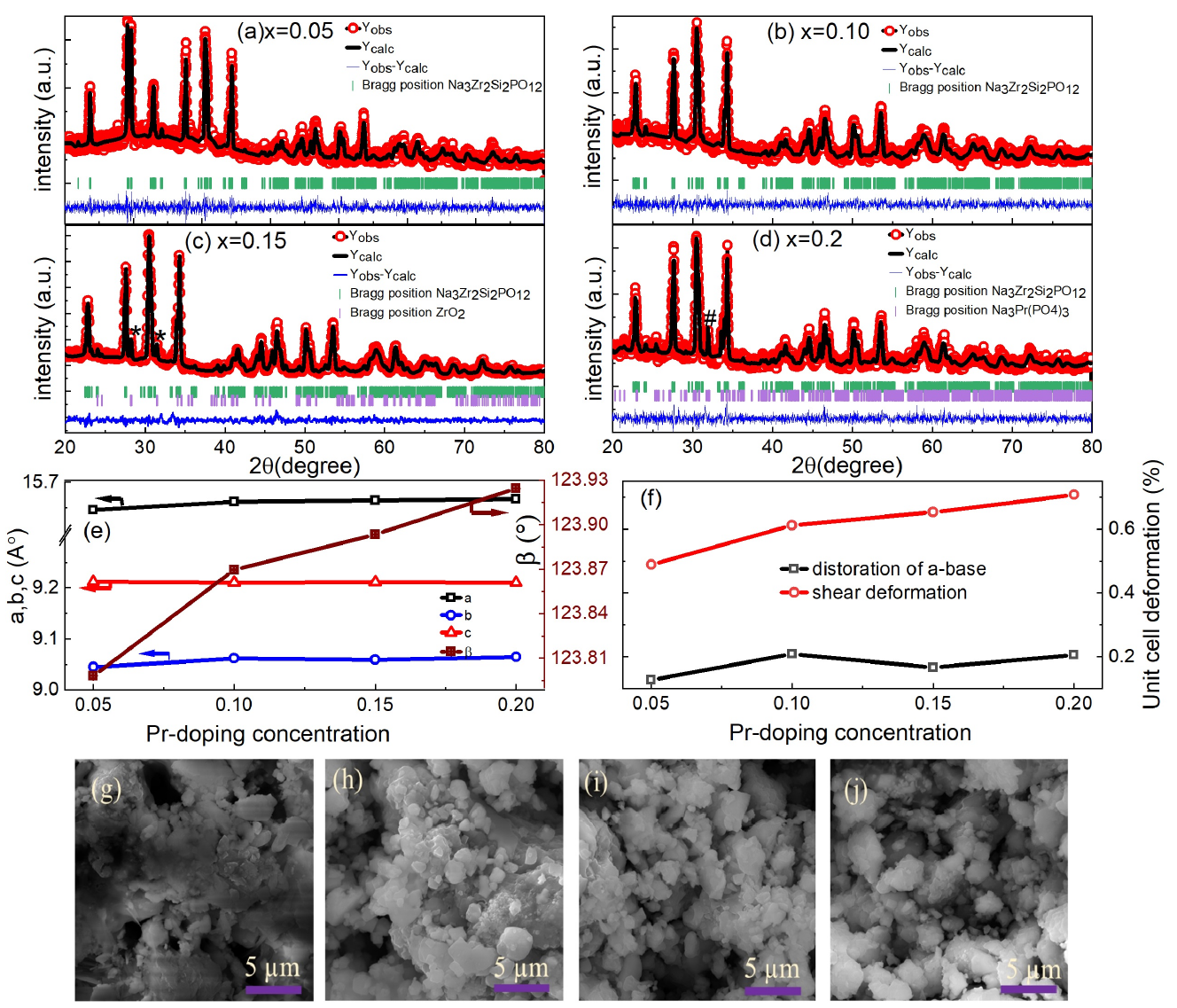}
    \caption{Reitveld refinement spectra of Pr-doped NASICON samples having the compositions of Na$_{3+x}$Zr$_{2-x}$Pr$_{x}$Si$_2$PO$_{\rm 12}$ $(x=0.05-0.2)$ are shown in (a-d). Here, open red symbols represent experimental data, and solid black line shows the Rietveld refined XRD patterns. The Bragg's positions correspond to the monoclinic NASICON phase is shown by green vertical lines and Bragg's peak corresponds to ZrO$_2$ and Na$_3$Pr(PO$_4$)$_3$ impurity phases are marked by * and \# symbols are shown as verticles lines of violet and wine color respectively. The blue color solid line represents the difference between observed and calculated data. (e) Lattice parameters of Pr-doped NASICON samples obtained after the refinement and (f) show the Distortion in a-base and shear deformation in unit cell as a function of Pr-doping concentration. (g-j) shows microstructural images of Na$_{3+x}$Zr$_{2-x}$Pr$_{x}$Si$_2$PO$_{\rm 12}$ $(x=0.05-0.2)$ samples obtained using field emission scanning electron microscope (FESEM). These images are taken at an acceleration voltage of 25 kV at a magnification of 5$\mu$m.}
\label{XRDSEM}
 \end{figure*}

\begin{figure*}
    \centering
    \includegraphics[width=1.0\textwidth]{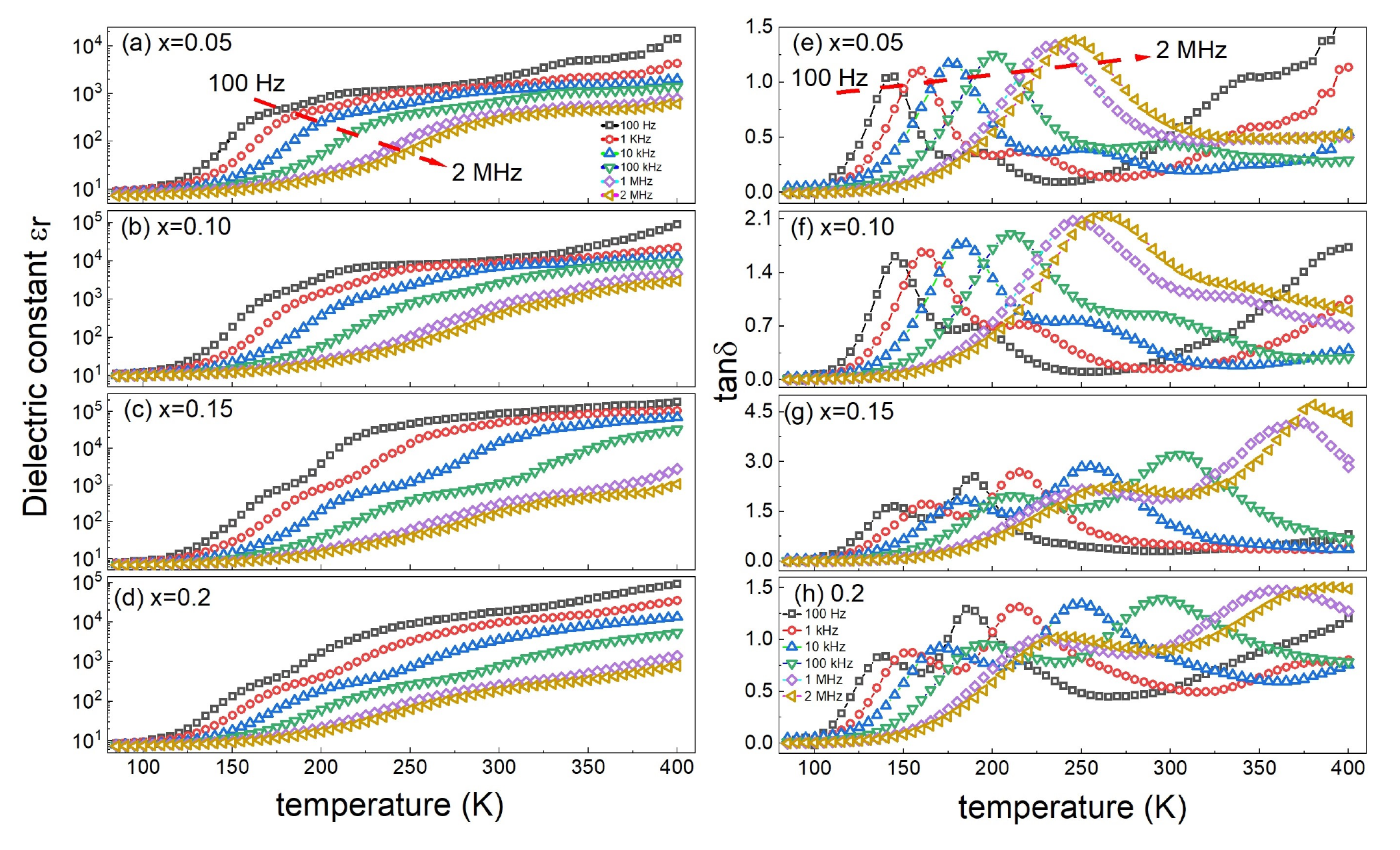}
    \caption{The temperature dependent variations in dielectric constant ($\epsilon_r$) (a--d) and loss tangent (tan$\delta$) (e--h) of Na$_{3+x}$Zr$_{2-x}$Pr$_{x}$Si$_2$PO$_{\rm 12}$ ($x=$ 0.05--0.2) samples at various selected frequencies from 100 Hz to 2 MHz. The arrows in panels (a) and (e) show the shifting of the relaxation peak toward high temperature side with an increase in the frequency.}
\label{CDT}
 \end{figure*}

The Rietveld refined room temperature XRD patterns of Na$_{3+x}$Zr$_{2-x}$Pr$_{x}$Si$_2$PO$_{\rm 12}$ ($x=$ 0.05--0.2) samples are shown in Figs.~\ref{XRDSEM}(a-d). The FullProf suite software \cite{Carvajal_CRNS_00} was used to analyze the XRD patterns by taking the pseudo-voigt peak shape and background correction is done using the linear interpolation. The refinement is proceeded by taking the scale factor, lattice parameters ($a$, $b$, $c$ and $\beta$), FWHM parameters ($u$, $v$, and $w$) and the shape parameters (Eta and $x$), the atomic positions ($x$, $y$, and $z$), asymmetry parameters and preferred orientation as free parameters. The occupancy parameters were kept fixed during the entire refinement process. The excellent agreement between the calculated and refined patterns shows the monoclinic phase formation having C2/c space group for all the samples having a small amount of ZrO$_2$ and Na$_3$Pr(PO$_4$)$_3$ impurity phase as denoted by the * and \# symbols, respectively. The reduced $\chi$$^2$ is achieved in the range of 1.8 to 2.2 for the best fit, and the refined lattice parameters are plotted in Fig.~\ref{XRDSEM}(e) and the values are in good agreement with \cite{Wang_JPCS_23, Jolley_JACS_15} as summarized in Table 1 of ref.~\cite{SI}. It has been found that $a$, $b$, and $\beta$ increase while $c$ decreases with an increase in Pr substitution, which indicate a tendency of phase transition from monoclinic to rhombohedral phase. To quantify the amount of unit cell deformation due to Pr substitution, the distortion in a--base and shear deformation are calculated using the following equations \cite{Wang_JPCS_23, Jolley_JACS_15}: 
\begin{subequations}
\begin{equation}
{\rm Distortion~of~a-base} = 1- \frac{a_M}{b_M \sqrt{3}}
\end{equation}
 \begin{equation}
{\rm Shear~deformation} = 1- \frac{3c_M~ {\rm cos}(180-\beta)}{a_M}
\end{equation}
\end{subequations}  
Here, a$_M$, b$_M$, c$_M$, and $\beta$ are the lattice parameters in the monoclinic NASICON phase where the distortion in a-base and shear deformation are used to determine the structural phase transition. The distortion graph with the amount of Pr substitution is shown in Fig.~\ref{XRDSEM}(f). It is observed that the shear deformation increases with $x$ due to larger ionic radius of the Pr ion (113 pm) as compared to the Zr ion (72 pm). The large ionic radii give rise to the unit cell volume expansion, resulting in shear deformation of unit cell. The Rietveld analysis shows that the $x=$ 0.2 sample contains Na$_3$Pr(PO$_4$)$_3$ impurity phase having a similar amount of Na and P ions, leading to variation in the Si/P ratio of the NASICON phase resulting in deformation of the lattice structure. The unit cell deformation with the amount of Pr substitution is shown in Fig.~\ref{XRDSEM}(f). The unit cell deformation increases with $x$ as addition of each Pr-ion require an additional Na ion inside the NASICON matrix, and larger radii of Pr ion as compared to Zr ion result in an increased unit cell volume, producing the distortion in the unit cell. 
Moreover, the microscopic images indicate the wide distribution of particles throughout the volume, as shown in Fig.~\ref{XRDSEM}(g--j). We find the agglomerated type of growth for $x=$ 0.05 sample, whereas the dense and compact microstructures are observed for all other samples. This type of microstructure is due to larger radii of Pr-ion, as higher concentration leads to an increase in unit cell volume, which gives the larger contact between grains. All the Pr doped samples show micro-cracks distributed along the grain boundaries, which may be related to the lattice thermal anisotropy during the sintering process. This lattice anisotropy increases the Si/P ratio due to the differential contraction along the $a$ and $c$ axes, reducing the contact between grains which result in the conduction loss and micro-cracks \cite{Wang_JPCS_23, Ma_JMCA_19}. The elemental composition is obtained using energy-dispersive x-ray (EDX) analysis, as shown in Fig.~1 of \cite{SI}, which confirms the required phase formation by the presence of constituent elements (Na, Zr, Pr, Si, P, O) in required stoichiometry. 

Figs.~\ref{CDT}(a--h) show the temperature dependent variation of electric permittivity (dielectric constant) ($\epsilon_r$) and dielectric loss tangent (tan$\delta$) of Pr-doped NASICON samples at selected frequencies between100 Hz and 2 MHz. It is observed that the electric permittivity (dielectric constant) increases with temperature and Pr-doping, reaching the magnitude of the order 10$^5$ at 400$^\circ$C (above 10\% doping) at 100 Hz, which is expected to be considered for high dielectric materials. The smaller values of electric permittivity at lower temperatures are due to the freezing of the charge carriers causing a delay in the carrier's polarization to the applied external field. The increasing behavior of electric permittivity with temperature is understood in terms of an increase in the polarization process due to the rapid movement of charge carriers as the available thermal energy increases with temperature. The dielectric constant variation shows a step-like behavior shifted towards  high-temperature side with an increase in frequency [as indicated by an arrow in $\epsilon_r$ data, see Fig.~\ref{CDT}(a)] indicating a relaxor-type behavior. The corresponding peaks in dielectric loss spectra (Figs.~\ref{CDT}(e--h)) also shifted towards the high-temperature side with an increase in frequency. Here we find the relaxation peak becomes broader with an increase in frequency. This type of relaxation phenomenon is explained using the interfacial or Maxwell-Wagner-Sillars (MWS) and space charge polarization models. According to these models a sharp increase in electric permittivity is due to the trapping of charge carriers inside the bulk part of the sample (MWS polarization) or accumulation of charge carriers at the interface of sample and electrodes (space charge polarization) \cite{Lu_JAP_06}. The frequency dependence of electric permittivity can be explained as follows: at lower values of frequency, the charge carriers build a potential barrier inside the matrix, leading to the piling of charge carriers at the grain boundary, resulting in a large dielectric constant. The lower values of the dielectric constant at higher frequencies are due to the reduction in space charge polarization, as the shorter period available with the carriers reduces the amount of polarization produced. The variations in electric permittivity with temperature show an increasing behavior due to enhanced mobility of charge carriers and provide an increased hopping due to additional thermal energy available at high-temperatures. This leads to additionally generated charge carriers that provide increased polarization, resulting in an increased dielectric constant. The dielectric constant is increased with doping due to reduced porosity with Pr doping, leading to higher contact between grains and an increase in the amount of defects. The additional number of carriers generated due to Pr doping reduces the vacant site available for Na migration and provides larger values of permittivity. The total contribution in dielectric constant at lower and higher frequencies is dominated by the grain boundary and grains \cite{Kumar_CI_24, Dar_JALCOM_15,Lin_PRB_05}. 

\begin{figure}
\includegraphics[width=3.4in]{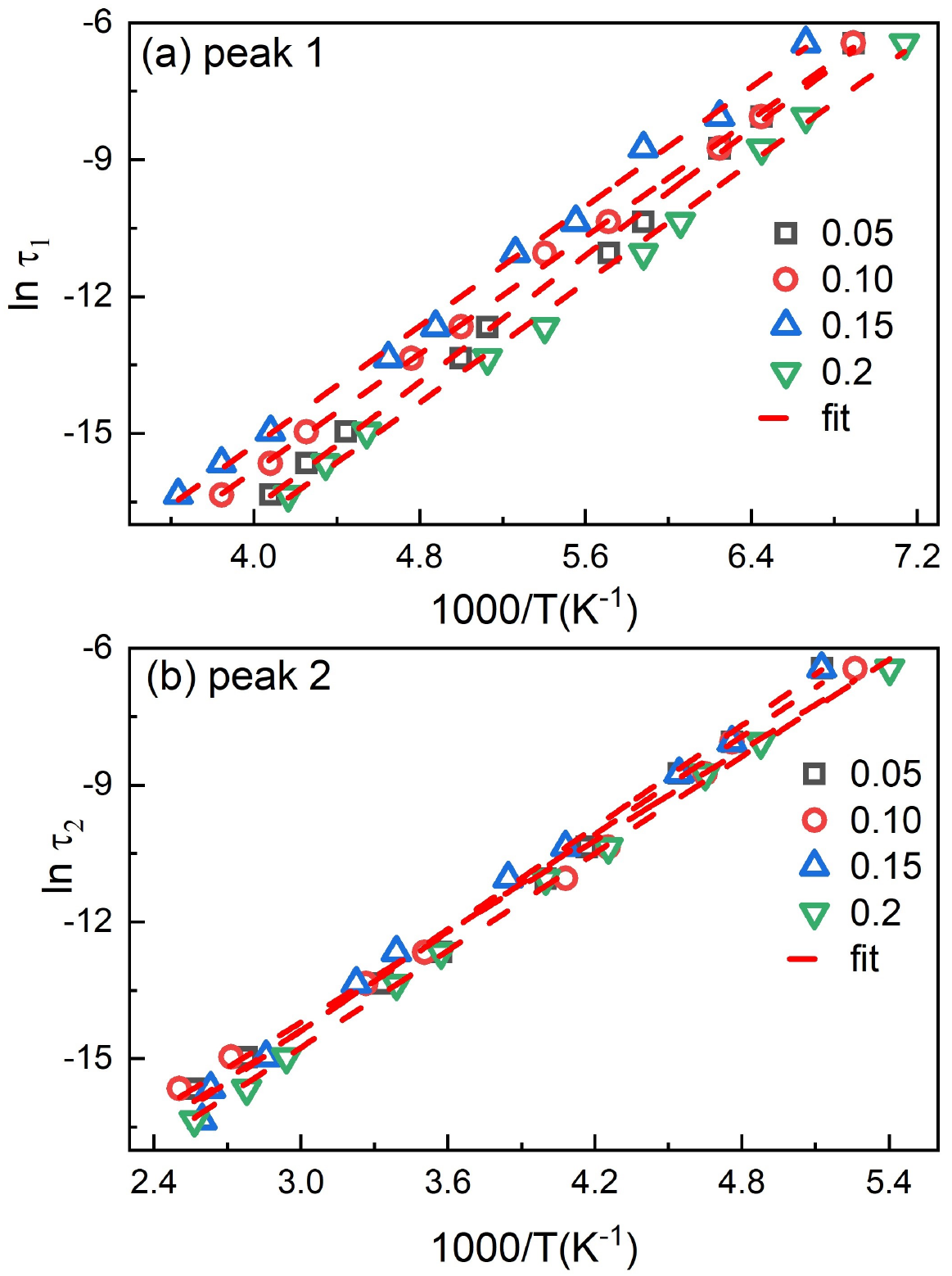}
\caption {The temperature dependence of relaxation time for peak 1 in (a) in the low-temperature range and for peak 2 in (b) at high temperatures. The open symbols show the measured data points and the solid lines show the linear fit.}
 \label{T1T2}
\end{figure}

\begin{figure*}
    \centering
    \includegraphics[width=1.0\textwidth]{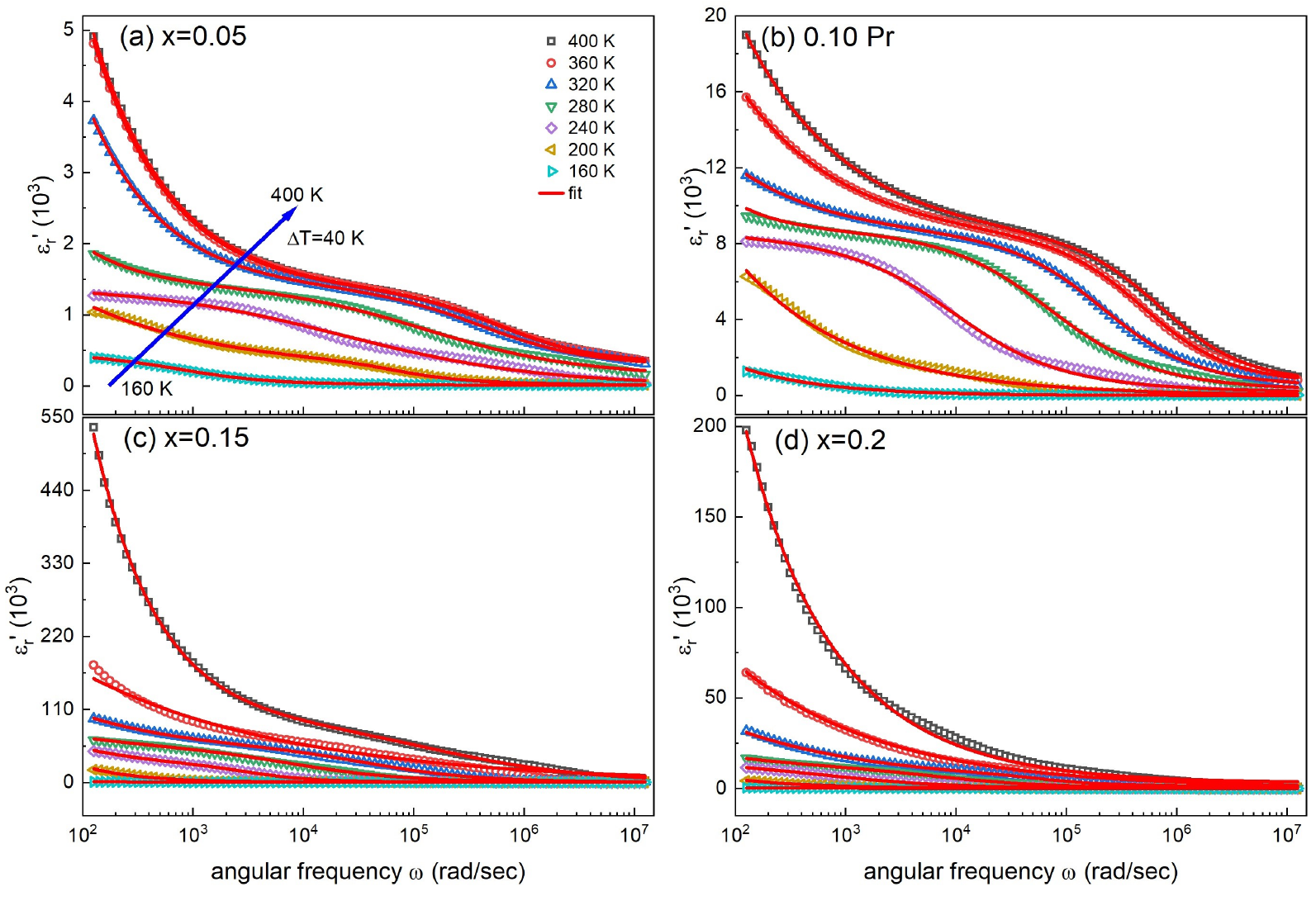}
    \caption{Isothermal frequency dependent real permittivity for Na$_{3+x}$Zr$_{2-x}$Pr$_{x}$Si$_2$PO$_{\rm 12}$ $(x=0.05-0.2)$ samples at various selected temperature. Here, open symbols are measured experimental data, and the solid line is fitting using the modified Cole-Cole relaxation, including the contributions from complex electrical conductivity.}
\label{RE}
 \end{figure*}

Note that the lag of polarization to the applied field appears as heat inside the material called dielectric loss. The dielectric loss in the measured frequency and temperature range comes from the combined effect of relaxation loss and conduction loss inside the material. The relaxation loss occurs during the polarization process in higher frequency regions once the induced polarization does not follow the applied external field, resulting in the generation of heat loss. In our experimental data, the height of the loss peak increases with frequency indicating the relaxation loss is dominating in Pr-doped samples at high frequency, see Figs.~\ref{CDT}(e--h). Also the conduction loss in the samples occurs due to their finite conductivity as well as through various types of defects, such as impurities, vacancies, point defects, and line defects at the grain boundary. At high temperatures, more charge carriers are generated due to large thermal energy available giving an increased joule heating in the material, hence larger dielectric loss \cite{Gao_CPB_19, Gao_JALCOM_19}. The peaks in the dielectric loss data are related to the hopping frequency of carriers; once the frequency of the applied electric field becomes equal to the hopping frequency we find a peak in the loss spectra. The relaxation peak becomes broad with an increase in frequency due to the spreading of relaxation time, as the frequency and relaxation time are related by the relation $\omega$$\tau$=1, where $\omega$ is the applied angular frequency and $\tau$ is associated relaxation time. The peak maxima of loss peak shifted towards higher temperature side with an increase in frequency, showing the Arrhenius type thermally activated relaxation. The temperature dependence of relaxation time is given by \cite{Ang_PRB_00, Jia_JAP_11, Meena_CI_22}
\begin{equation}
     \tau = \tau_0 ~ exp(\frac{E_{a}}{k_B T_m})
\label{tau}
 \end{equation}
here $\tau_0$ is the pre-exponential factor called relaxation time at infinite temperature has the order of atom vibrational period ($10^{-13}$ sec), $E_{a}$ is the activation energy of dipolar relaxation, $\tau$ is calculated relaxation time, k$_B$ is Boltzman constant and $T_m$ is the peak temperature of the loss peak. The activation energy of relaxation peaks is calculated using equation~\ref{tau} (by taking the slope of ln $\tau$ vs $\frac{1000}{T}$ graphs), which shows the activation energies for the peak 1 between 0.28 eV to 0.30 eV, while activation energies for high-temperature relaxation peak varies between 0.29 to 0.32 eV, as shown in Figs.~\ref{T1T2}(a) and (b), respectively. We find that the magnitude of relaxation activation energy is nearly the same for the both peaks, indicating a similar type of relaxation over the measured temperature range. The characteristic relaxation time $\tau_0$ is obtained from the intercept of the curve, which found to vary in the range of 1.2--4.3 $\times$ $10^{-13}$ sec. 

\begin{figure*}
    \centering
    \includegraphics[width=0.97\textwidth]{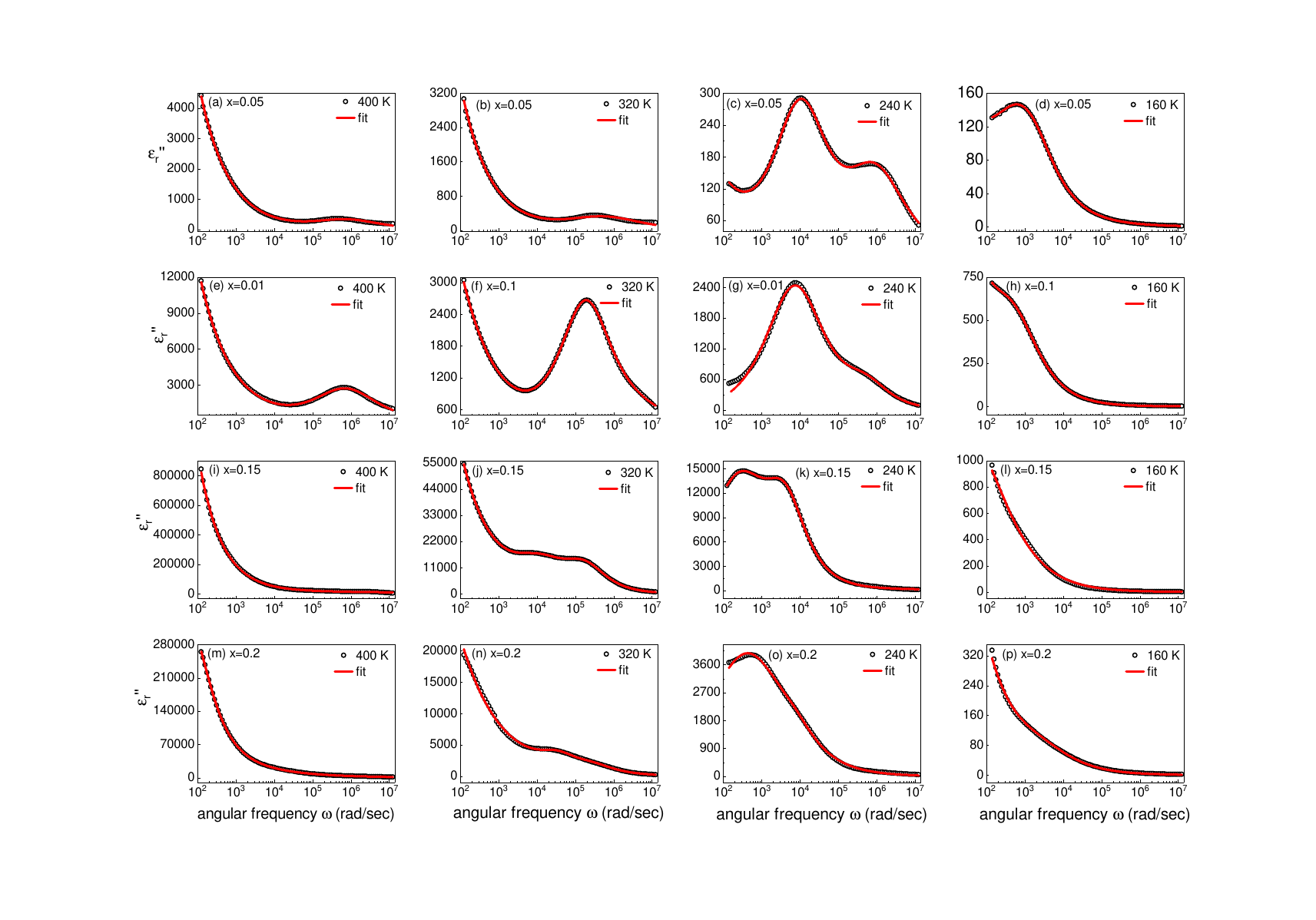}
    \caption{Isothermal frequency dependent imaginary part of electric permittivity ($\epsilon$ $^{''}$) variations with frequency at selected temperatures of Na$_{3+x}$Zr$_{2-x}$Pr$_{x}$Si$_2$PO$_{\rm 12}$ $(x=0.05-0.2)$ samples. Here, the open symbols show the measured experimental data, and the solid line is the fitting using modified Cole-Cole relaxation.}
    \label{IE}
\end{figure*}

To further understand the higher values of electric permittivity and associated relaxation phenomena, the real ($\epsilon$ $^{'}$) and imaginary ($\epsilon$ $^{''}$) parts are plotted as a function of frequency, as shown in Figs.~\ref{RE} and \ref{IE}, respectively, at selected temperatures. We observe that the dielectric constant increases with temperature and follows the inverse behavior with frequency. The real part of permittivity ($\epsilon$ $^{'}$) varies with frequency showing a non-Debye-like relaxation with a step-like decrease and shifting of the relaxation peak towards high frequency with an increase in temperature. The larger values of permittivity at lower frequencies are due to an increase in charge carrier concentration at the interface, resulting in enhanced polarization making these materials a good choice for low-frequency energy storage applications \cite{Peng_CI_20}. The carrier's polarization process at higher frequencies was so rapid that no charge accumulation at the interface resulted in decreased permittivity. This type of frequency-dependent variation can be explained using the space charge (interfacial) polarization mechanism; according to these models, the charge carriers inside the materials are impeded by the grain boundary, which helps to prevent charge migration. In this case, the charge carriers are poles up at the grain boundaries producing a local polarization within the grains \cite{Pawlicka_EA_19, Thongbai_JPCM_08}. The temperature dependence of permittivity can be explained as the density of the carriers contributing to the polarization process increasing with temperature, leading to an increase in space charge polarization, resulting in increased permittivity. Notably the space-charge polarization is sensitive up to 10$^4$ rad/sec giving the larger values of electric permittivity. The decrease in the dielectric constant at high-frequency values is due to the random orientation of carriers responsible for dielectric phenomena. The strong dispersion behavior of electric permittivity is due to the thermally activated charge carriers like space charge carriers, defects, and their related complex phenomena \cite{Rehman_JAP_15}. In general, there are two main effects responsible for the polarization; the frequency of the applied field and measurement temperature. Firstly we consider the effect of frequency, at a fixed temperature, for lower values of frequency the charge carriers follow the applied field to produce larger values of polarization resulting in higher values of permittivity, while at higher frequencies the field variation is so rapid, the charge carriers are no longer be able to follow the applied field resulting in lower values of permittivity. To consider the effect of temperature at fixed frequency, at low temperature the freezing of dipoles gives the decay in the polarization process concerning the applied field, resulting in decreased values of electric permittivity. At higher temperatures, the rate of polarization increases due to the large thermal energy available for carriers giving the higher values of electric permittivity with a shifting of relaxation peak towards higher frequencies \cite{Lin_PRB_05, Thongbai_JPCM_08}.

Now we study the frequency-dependent behavior of real ($\epsilon$$^{'}$) and imaginary ($\epsilon$$^{''}$) part of permittivity, which can be explained using the modified Cole-Cole equation including the conductivity term given by \cite{Abdelkafi_JAP_06} 
\begin{equation}
    \epsilon^*=\epsilon^{'} - \iota \epsilon^{''}=\epsilon_{\infty} + \frac {\epsilon_s-\epsilon_{\infty}}{[1+ (\iota \omega \tau)^{\alpha}]} + \frac{\sigma^{*}}{\iota \omega^s}
\label{epsilon}
\end{equation}
here, the $\epsilon_s$, $\epsilon$$_\infty$ are the low and high-frequency limit of the permittivity, respectively, $\omega$ is the applied angular frequency, $\tau$ is the characteristics relaxation time, and $\alpha$ are the shape parameters represents the deviation from ideal Debye type relaxation having the values between 0 to 1. For ideal Debye type relaxation (considering no interaction among the dipoles or charge carriers) $\alpha$=0. The values of $\alpha$ greater than zero (considering the significant interaction among the carriers) indicate the distribution of relaxation times, which leads to a broad peak shape rather than a single Debye peak. The $\sigma$$^*$ is complex conductivity given by $\sigma$$^*$ = $\sigma_1$ + $\iota$ $\sigma_2$. Here, the $\sigma_1$ represents the conductivity due to free charge carriers and $\sigma_2$ is the conductivity due to localized or bound charge carriers. The value of $s=$ 1 shows the ideal ohmic-type behavior of the complex conductivity, $s<1$ indicates the distribution of the polarization mechanism. Using equation \ref{epsilon}, the real and imaginary parts of permittivity can be written as \cite{Abdelkafi_JAP_06, Thongbai_JPCM_08}:  
\begin{subequations}
\begin{equation}
\epsilon^{'}= \epsilon_{\infty}+ \frac{(\epsilon_s-\epsilon_\infty)[1+(\omega\tau)^{1-\alpha} sin(\frac{\alpha \pi}{2})]}{[1+2 (\omega\tau)^{1-\alpha}  sin(\frac{\alpha \pi}{2})+ (\omega\tau)^{2-2\alpha}]}+\frac{\sigma_2}{\epsilon_0 \omega^s}
\label{E'}
 \end{equation}  
\begin{equation}
\epsilon^{''}= \frac{(\epsilon_s-\epsilon_\infty)[(\omega\tau)^{1-\alpha} cos(\frac{\alpha \pi}{2})]}{[1+2 (\omega\tau)^{1-\alpha}  sin(\frac{\alpha \pi}{2})+ (\omega\tau)^{2-2\alpha}]}+\frac{\sigma_1}{\epsilon_0 \omega^s}  
\label{E''}
 \end{equation}  
\end{subequations}
From equations \ref{E'} and \ref{E''}, we can see that the localized defects sites and interfaces contribute to the electric permittivity due to bound charge carriers ($\sigma_2$), while the free charge carriers contributes in the dielectric loss ($\sigma_1$). The first term in the dielectric loss equation (\ref{E''}) gives the contribution of permanent dipoles following the short-range mobility of carriers, while the second term gives the contribution from free carriers possessing the long-range migration of carriers ($d.c.$ conductivity part). The real and imaginary parts of permittivity are fitted using the equations \ref{E'} and \ref{E''}. The fitted curves are shown in Figs.~\ref{RE} and \ref{IE}, respectively. We find that the imaginary permittivity data exhibit larger values of permittivity at higher temperatures and lower frequencies, which are due to the free carrier motion giving large interfacial polarization at the grain boundary. This type of behavior is explained using the Maxwell–Wagner theory and Koop’s model; according to these models, the overall dielectric response is the combined effect of the poorly conducting grain boundaries separated by well-conducting grains \cite{Javed_MRB_23, Pawlicka_EA_19}. The ($\epsilon$$^{''}$) curves exhibit lower values of permittivity at lower temperatures due to the reduction in the anharmonic lattice force between the crystal lattice and phonons in the crystal lattice. The larger values at higher temperatures are due to an increase in thermally enhanced electrical conduction. The imaginary part of permittivity ($\epsilon$$^{''}$) shows the double relaxation peaks at lower temperatures, and the height of these peaks increases with doping due to matching of applied field frequency and reduced mobility of charge carriers \cite{Rehman_JAP_15, Abdelkafi_JAP_06, Javed_MRB_23}. 

\begin{figure*}
    \centering
    \includegraphics[width=1.0\textwidth]{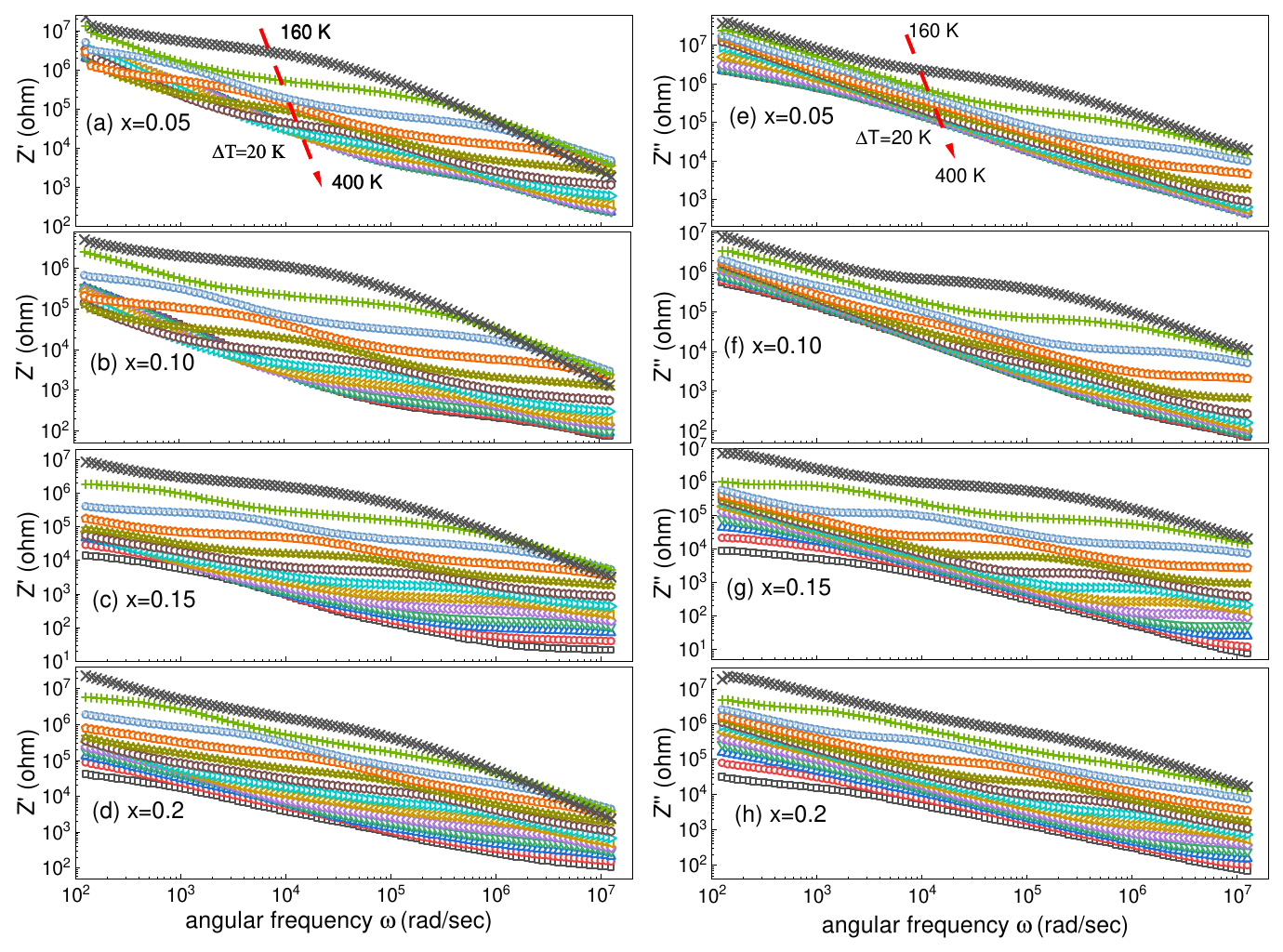}
    \caption{The real ($Z'$) and imaginary ($Z''$) part of total impedance are shown in figure (a-d) and (e-h) as a function of frequency for Na$_{3+x}$Zr$_{2-x}$Pr$_{x}$Si$_2$PO$_{\rm 12}$ $(x=0.05-0.2)$ (a-d) samples at selected temperatures. The arrow indicates the direction of increasing temperature.}
    \label{Z'Z''}
\end{figure*}

 In order to understand the relaxation processes and conduction mechanisms that contribute to the dielectric phenomena as well as the role of microstructure, we can use an equivalent circuit model for analyzing the resistive and reactive response at grain and interface of polycrystalline materials. Each contribution can be modeled using the parallel combination of resistor (R) and capacitor (C), and all the components are connected in series corresponding to grain, grain-boundary, and electrode contribution \cite{Singh_JAP_12, Javed_MRB_23, Schmidt_PRB_09}. The real and imaginary parts of normalized impedance data are shown in Figs.~\ref{Z'Z''}(a--d) and \ref{Z'Z''}(e-h), respectively, as a function of frequency in the log-log scale. The impedance data are normalized using the geometrical factor ($\frac{A}{2d}$), where $A$ is the sample electrode area and $d$ is the thickness of the pellet \cite{Chandrasekhar_JPCM_12, Schmidt_PRB_09}. The real part of impedance ($Z^{'}$) represents the opposition of alternating current and its magnitude is found to be nearly the same for all studied Pr-doped samples. The magnitude of $Z^{'}$ is dominated by the highest resistive component inclined towards high frequency, which decreases with temperature, i.e., the negative temperature coefficient of resistance, which indicate the semiconducting/insulating type behavior in the measured temperature range. The decrease in impedance can be due to a reduction in barrier heights, which enhanced the mobility of charge carriers by decreasing the activation energy. We observe one broad relaxation peak towards the higher frequency side at lower temperatures, and another relaxation peak appears at the lower frequency side, which shifts towards higher frequency with an increase in temperature, as marked by solid and dashed arrows, respectively, in Figs.~\ref{Z'Z''}(a--d). These two clear broad peaks were found in all the Pr-doped samples and the intensity found to decrease with increasing the temperature. The broad nature of relaxation peaks shows the non-Debye nature due to the microscopic contributions of various micro-constituents where the resistive and reactive components may play a crucial role in the relaxation \cite{Abdelkafi_JAP_06, Javed_MRB_23}. The frequency dependence of real impedance can be divided into three categories: (a) the nearly constant values at lower frequencies represent the d.c. contribution, (b) the rapid decrease at intermediate frequencies shows ac conduction inside the material, and (c) the merging of all the curves at higher frequencies is due to the possible release of space charge at the grain boundary interfaces due to a reduction in barrier heights \cite{Zhang_MRB_22, Javed_MSEB_21, Dhiman_SI_23}. 

\begin{figure*}
    \centering
    \includegraphics[width=1.0\textwidth]{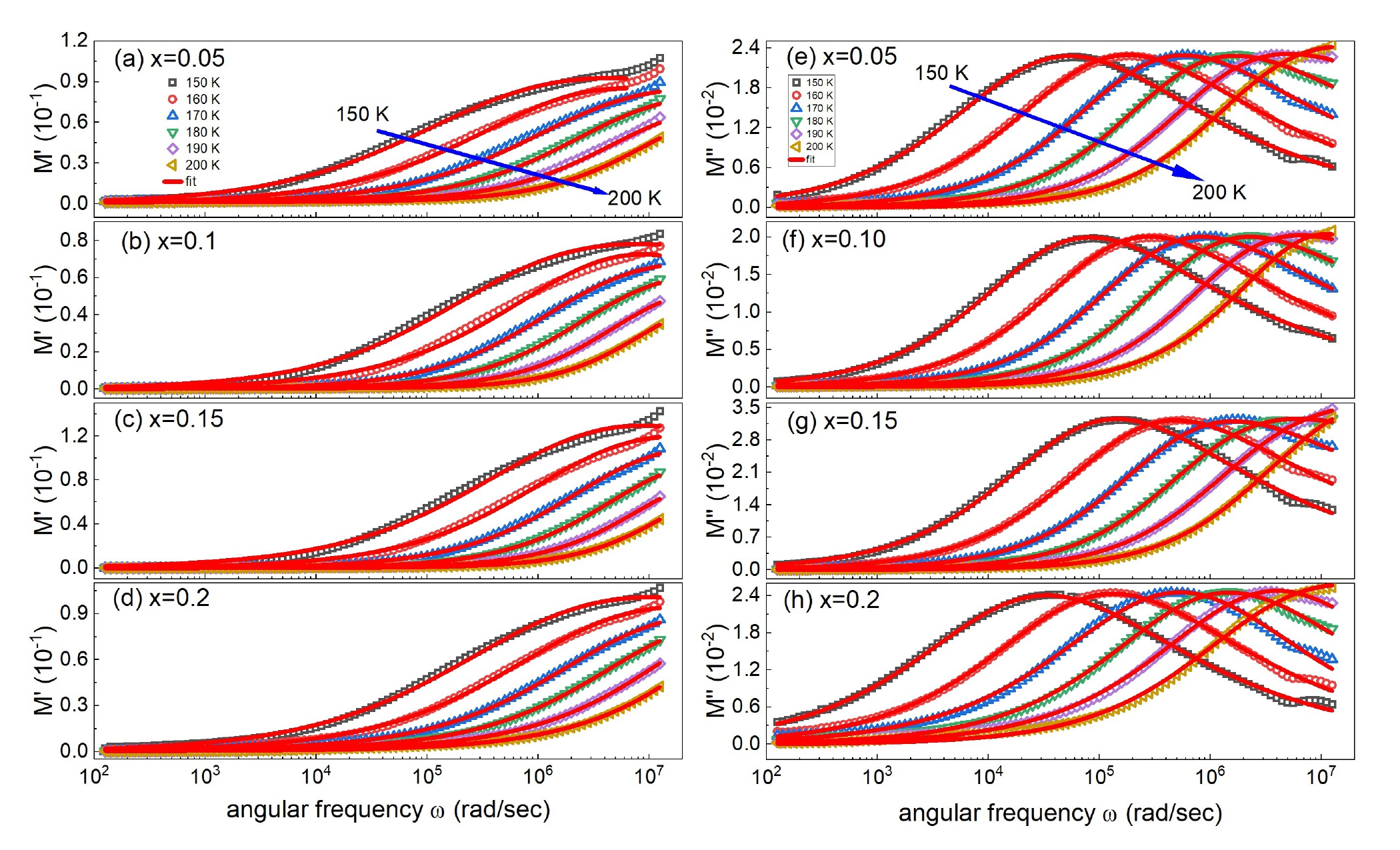}
    \caption{The variation of real ($M^{'}$) (a-d) and imaginary ($M^{''}$) (e-f) part of the electric modulus function at selected temperature as a function of frequency. Here, open symbols represent the measured experimental data, and a solid line is fit using the electric modulus function. The $M^{'}$ and $M^{''}$ above 200 K are shown in Figs.~3 and 4 of ref.~\cite{SI}.}
\label{MMT}
 \end{figure*}

The imaginary ($Z^{''}$) part of impedance study [as shown in Figs.~\ref{Z'Z''}(e-h)] measures the phase lag between the applied voltage and the measured current where the  position of relaxation peaks and analysis provide information about various contributions like from electrodes, grain, and grain boundaries present in polycrystalline samples \cite{Kumar_CI_24, Liu_PRB_04}. Here, all the Pr-doped samples show the relaxation peaks at the higher frequency side, which are attributed to the contribution from grains or bulk. A weaker relaxation peak at lower frequencies for high temperatures is attributed to grain-boundary relaxation, which represents the movement of charge carriers at the interface of different grains. The relaxation time ($\tau$) related to the grain boundary is much larger (10$^{2}$ times) than the grains; therefore the grain boundary relaxation happens at a much lower frequency and lower the mobility of carriers as compared to the grains. This is consistent as the resistance at the grain boundary is larger than the grains or bulk due to the trapping of charge carriers at the interface, oxygen vacancies,  and defects present at the interface or boundaries in the polycrystalline samples \cite{Gao_JALCOM_19, Gao_CPB_19}. The peak frequency of the impedance spectra is related by $\frac{1}{2 \pi RC}$, the peak position for grain boundary is at much lower frequencies as compared to the grains due to the larger resistance and capacitance associated with the boundaries. It is observed that the relaxation peaks shift towards higher frequency side with an increase in temperature, suggesting the temperature-dependent activation of relaxation with the distribution of relaxation times and a decreasing intensity pattern. These suggest that the two relaxations (grain and grain boundary) are of non-Debye nature over the measured frequency and temperature range \cite{Nasari_CI_16, Taher_MRB_16}. In order to understand the charge carrier dynamics during the relaxation process, the combined analysis of impedance spectra and electric modulus is required \cite{Rehman_JAP_15,  Kaswan_JALCOM_21}. Here, the impedance plot is dominated by highest resistive element and modulus spectra are dominated by the lowest capacitive element, as shown in Fig.~2 of ref.~\cite{SI}. As the peaks related to impedance (Z'') and modulus (M'') do not match each other confirming the localized short-range mobility of charge carriers. The higher grain boundary resistance, as discussed above, is responsible for the short-range mobility of charge carriers by the hopping type conduction mechanism \cite{Xue_JALCOM_24, Kumar_MCP_13, Kaswan_JALCOM_21}. 

Now we present the electric modulus study to determine the different components of the relaxation behavior such as ion/carrier hopping rate, and its conduction mechanism under the influence of applied electric field \cite{Moynihan_PCG_73, Macedo_PCG_1973}. Through the detailed analysis, we can understand different contributions in total electrical conduction by identifying equivalent resistance and unequal capacitance contributions of grains, grain boundary, and electrode polarization effects \cite{Singh_JALCOM_17, Javed_MRB_23, Moynihan_PCG_73, Macedo_PCG_1973}. The complex electric modulus $M^*$=$M^{'}$ +$i$ $M^{''}$ is represented by Fourier transformation of relaxation function $\phi$$(t)$, as given by \cite{Javed_MSEB_21,Pal_JAP_19}
\begin{equation}
M^{*} (\omega)= M_\infty \Biggr [1-  \int_{0}^{\infty} exp(-i\omega t) (\frac{-d\phi}{dt}) \,dt  \Biggr]
 \end{equation}
where the M$_{\infty}$ = $\frac{1}{\epsilon_{\infty}}$ is the high frequency asymptotic value of the real modulus $M^{'}$$(\omega)$. The $\phi$$(t)$ is the relaxation function that represents the decay of the electric field inside the ionic conductor measured in the time domain that arises due to the movement of charge carriers. For ideal Debye case, relaxation function $\phi(t)$ is exponential in nature, while for non Debye type the $\phi(t)$ is approximated by the well known Kohlrausch-Williams-Watts (KWW) decay function, as given by
\begin{equation}
\phi (t)= \Biggr [ exp -(\frac{t}{\tau})^{\beta} \Biggr]
\end{equation}
here, the $\tau$ is conductivity relaxation time, and $\beta$ is a stretched exponent called the Kohlrausch exponent containing the values between 0 to 1, which signifies the distribution in relaxation times due to strong correlation between structural disorder and charge carrier dynamics. For the ideal Debye type relaxation, the value of $\beta$ should be unity, the smaller values of $\beta$ indicate the larger deviation from the ideal Debye relaxation. 
\begin{figure}[h] 
\includegraphics[width=3.5in]{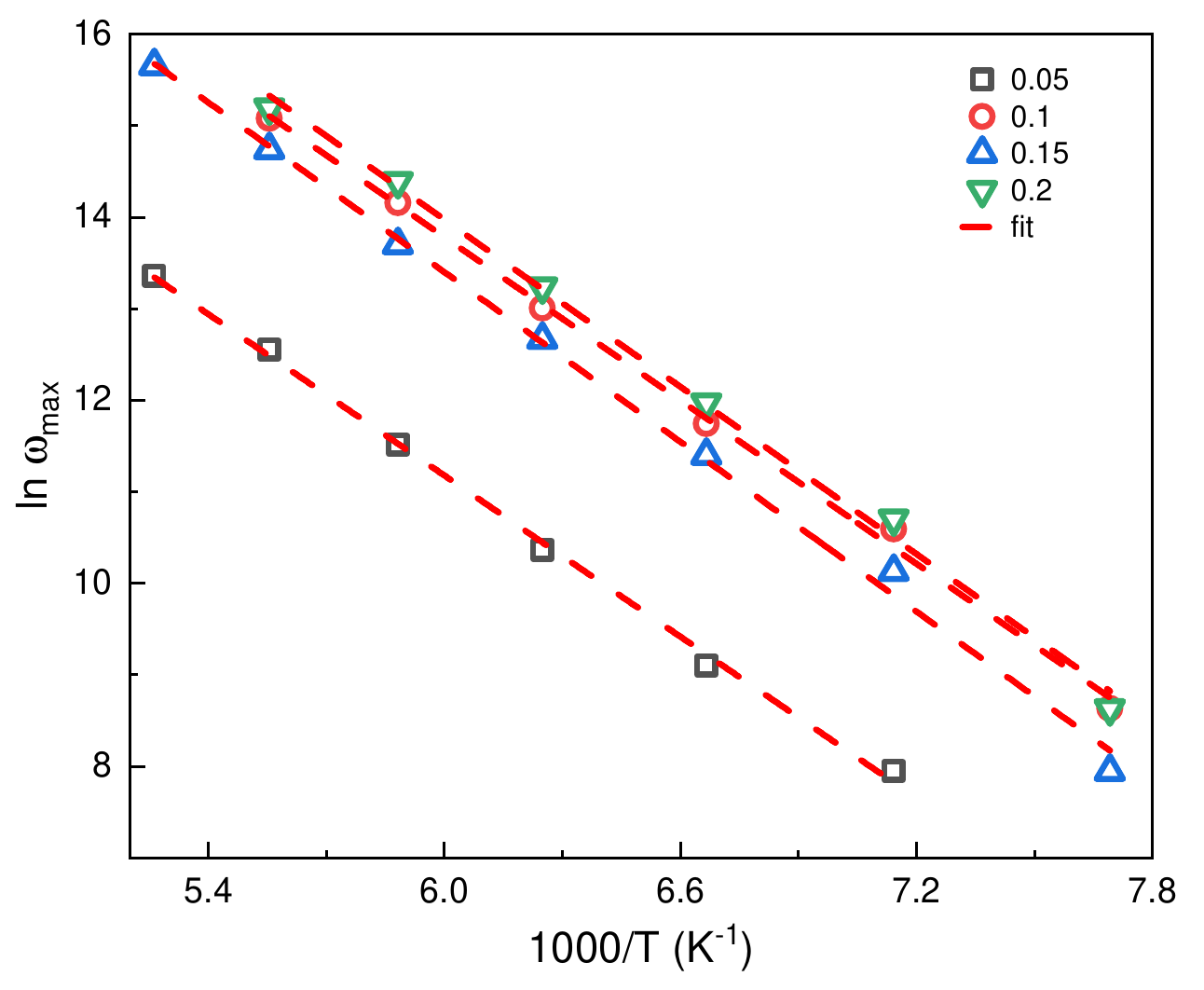}
\caption{Arrhenius plot of peak modulus frequency with temperature for Pr-doped NASICON samples. Here, open symbols represent the measured peak relaxation frequency, and the solid line represents the Arrhenius linear fit.}  
\label{EM}
\end{figure}
The temperature-dependent variation of the real part of the electric modulus function ($M^{'}$) is shown in Figs.~\ref{MMT}(a-d) as a function of frequency at selected temperatures. It is found that the magnitude of ($M^{'}$) approaches zero value at lower frequencies for all the measured temperatures indicating the absence of any electrode polarization effect in the relaxation and conduction process. The lower values of ($M^{'}$) at smaller frequency are explained by the lack of force to the available carriers under the influence of the applied field. The ($M^{'}$) spectra show an increase in magnitude with an increase in frequency having the asymptotic nature approaches towards M$_{\infty}$ for higher values of frequency called the high-frequency limit of ($M^{'}$). This type of variation in ($M^{'}$) spectra shows the capacitive nature of our samples \cite{Singh_JALCOM_17}. The magnitude of ($M^{'}$) decreases with an increase in temperature due to the increased mobility of carriers, as the carriers have sufficient energy to get oriented towards the applied field at higher temperatures. The dispersive region shifted towards higher frequencies with an increase in temperature giving the possibility of long-range movement of charge carriers above a certain frequency \cite{Pal_JAP_19, Sondarva_JALCOM_21}. This type of dispersive nature of ($M^{'}$) function suggests the short-range mobility of charge carriers in the conduction process and gives the spread of relaxation times with an increase in temperature \cite{Javed_MSEB_21, Deb_JAP_10, Nasari_CI_16}.  

In Figs.~\ref{MMT}(e-h), the frequency dependence of the imaginary part of the electric modulus ($M^{''}$) is presented at selected temperatures. The relaxation peak shifts towards higher frequency side, confirming the ionic nature of studied samples having faster ion dynamics with decreased relaxation time with an increase in temperature. The large values of relaxation frequency at high temperatures give the decreased relaxation time due to enhanced mobility of charge carriers. The peak frequency of the spectra $\omega_{max}$ differentiates the charge carriers' conduction mechanism. The conduction below the peak frequency is governed by the long-range movement of charge carriers, and above the peak frequency, it is related to the short-range mobility of charge carriers where the charge carriers are confined over a potential well, making the localized motion contributes to the conduction process. It is found that the imaginary modulus ($M^{''}$) plots are asymmetric and the the relaxation peaks skewed towards high frequency, which indicate the non-Debye-type relaxation for all the samples  \cite{Deb_JAP_10, Nasari_CI_16}. The electric modulus ($M$) and its real ($M'$) and imaginary ($M''$) parts are analyzed using the Havriliak-Negami (HN) model with Kohlrausch-Williams-Watts (KWW) approach to understanding the non-Debye relaxation dynamics; as given by the following equation:\cite{Nasari_CI_16, Pal_JAP_19}
\begin{equation}
M^*(\omega)=M_{\infty} + \frac{(M_{s}-M_{\infty})}{[1+(i\omega \tau)^{\alpha}]^{\gamma}}
\label{M}
 \end{equation}
where the M$_{s}$ and M$_{\infty}$ are low and high-frequency limit values of electric modulus, respectively, the $\alpha$ and $\gamma$ are the shape parameters characterize the symmetric nature of the complex electric modulus having the values smaller than one, and $\tau$ is the relaxation time. The real and imaginary parts of the modulus are presented using the below equations: \cite{Islam_MRB_23, Alvarez_PRB_91}
\begin{subequations}
\begin{equation}
M^{'}=M_{\infty} + \frac{(M_{s}-M_{\infty}) ~ Cos(\gamma \phi)}{[1+2(\omega\tau)^{\alpha} ~ Cos (\frac{\pi \alpha}{2})+(\omega\tau)^{2\alpha}]^{\frac{\gamma}{2}}}
\label{M'}
 \end{equation}  
\begin{equation}
M^{''}= \frac{(M_{s}-M_{\infty}) ~ Sin(\gamma \phi)}{[1+2(\omega\tau)^{\alpha} ~ Cos (\frac{\pi \alpha}{2})+(\omega\tau)^{2\alpha}]^{\frac{\gamma}{2}}}
\label{M''}
 \end{equation}  
\end{subequations}
 where 
\begin{equation}
\phi= \arctan [\frac{(\omega\tau)^{\alpha} ~ Sin (\frac{\pi \alpha}{2})}{1+(\omega\tau)^{\alpha} ~ Cos (\frac{\pi \alpha}{2})}]
\end{equation}
The real and imaginary parts are fitted using equations \ref{M'} and \ref{M''}, respectively, as shown in Fig.~\ref{MMT} considering the M$_{s}$, M$_{\infty}$, $\tau$, $\alpha$ and $\gamma$ as fitting parameters. The values of $\alpha$ and $\gamma$ are found to be in the range of 0.5--0.98 and 0.16--0.96, respectively, which are less than 1 for all the samples, suggesting the non-Debye type relaxation. This may be due to the defects present in the samples and non-exponential behavior is due to the charge ordering or electric polarization generated due to the background electric field \cite{Islam_MRB_23, Alvarez_PRB_91}. 

       \begin{figure}[h]
    \centering
    \includegraphics[width=0.48\textwidth]{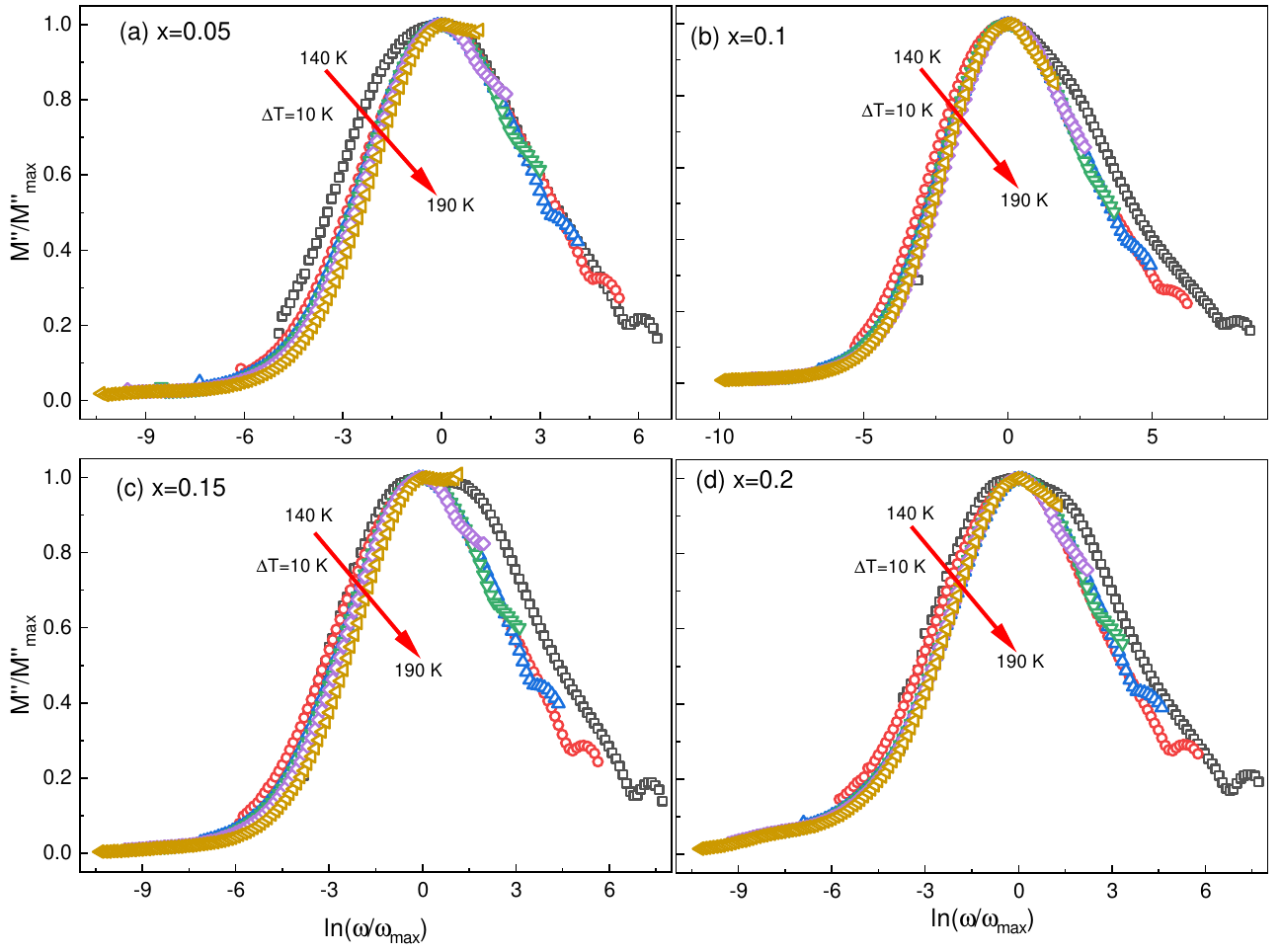}
    \caption{The scaling analysis spectra of imaginary modulus ($M^{''}$) for Pr-doped NASICON samples.}
\label{MS}
 \end{figure}

          \begin{figure*}
    \centering
    \includegraphics[width=0.95\textwidth]{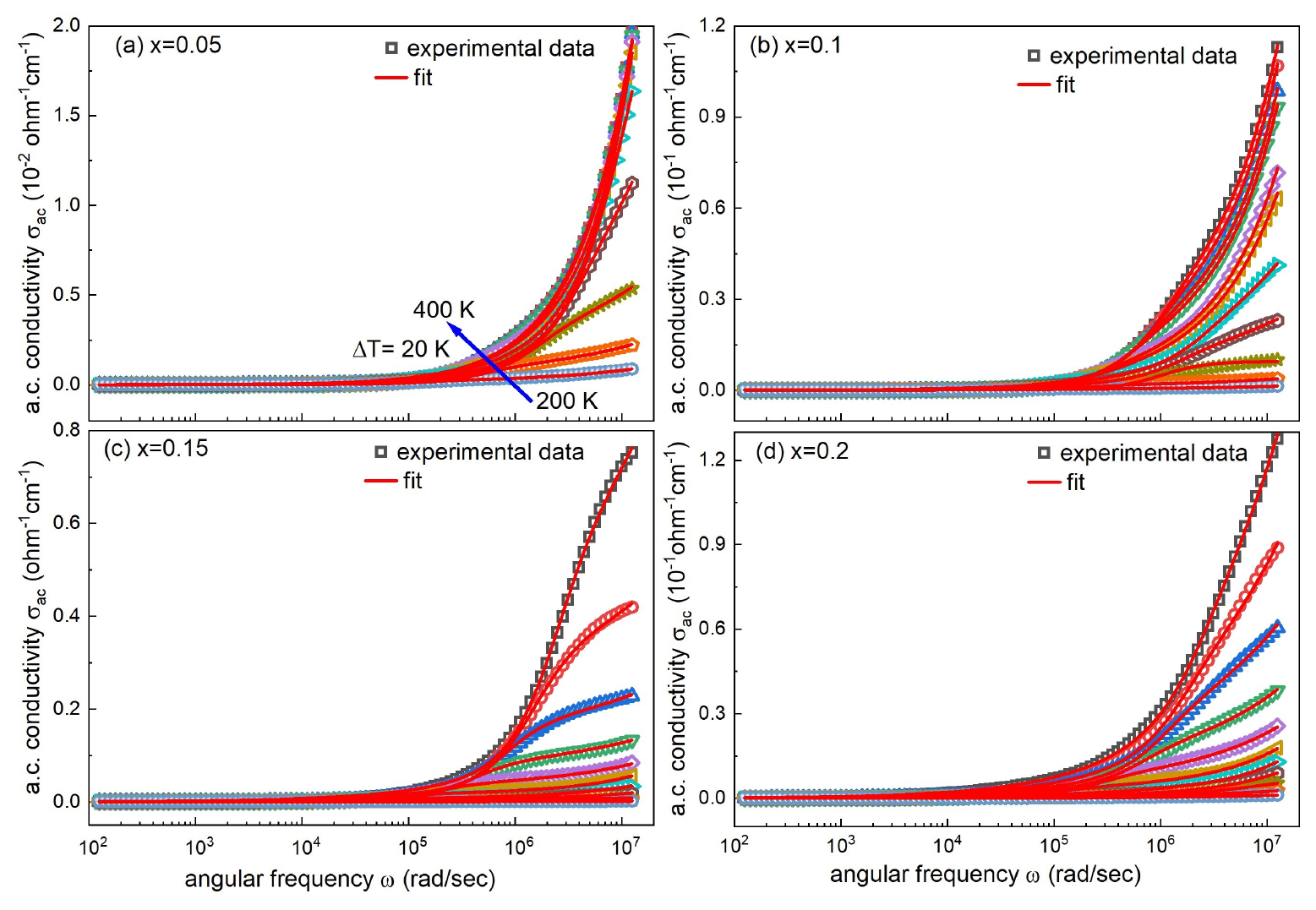}
    \caption{The frequency dependence of a.c. conductivity at selected temperatures for Pr-doped samples. Here, the open symbol represents the measured experimental data, and open red solid line shows the fit using the modified power law.}
\label{ACS}
 \end{figure*}

We find that the frequency related to the peak in imaginary modulus (called peak frequency $\omega_{max}$) shifts towards the high-frequency side with an increase in temperature and showing the Arrhenius-type thermal activation. The activation energy of the peak relaxation frequency is determined using the relation below: \cite{Javed_MRB_23, Kumar_CI_24, Nasari_CI_16}
\begin{equation}
    \label{omega-Arr}
    \omega_{m} = \omega_0 ~ exp (\frac{-E_M}{k_B T})
\end{equation}
where $\omega_0$ is the pre-exponential factor, k$_B$ is the Boltzmann constant, T is the measured temperature and $E_M$ is the activation energy of the relaxation. The activation energy values of relaxation are determined using equation \ref{omega-Arr}, which found to be 0.26$\pm$0.01, as shown in Fig.~\ref{EM}, which indicate a similar type of thermal activation over the measured temperature range for all the samples. The magnitude of activation energy suggests the polaron relaxation is responsible for charge conduction. Further, the scaling analysis at selected temperatures is shown in Figs.~\ref{MS}(a-d) where each angular frequency is scaled by the peak angular frequency ($\omega_{max}$), and imaginary modulus ($M^{''}$) is scaled by the peak modulus value ($M_{max}^{''}$). Notably, all the curves merge over the measured temperature range showing similar type of relaxation, which suggests the temperature independent dynamic process in all the studied samples \cite{Deb_JAP_10, Hadded_RSC_20, Kaswan_JALCOM_21}.

Finally we present the analysis of {\it a.c.} conductivity ($\sigma$) to understand the effect of relaxation, localized hopping, and diffusion of charge carriers in the vicinity of the Fermi level. The $\sigma$ depends on the geometry of the sample and calculated using the relation $\sigma$($\omega$)=$\omega$$\epsilon_0$$\epsilon_r$tan$\delta$, here $\epsilon_r$ is relative electric permittivity (dielectric constant), $\epsilon_0$ is the permittivity of free space, tan$\delta$ is the dielectric loss and $\omega$ is the angular frequency of the applied field. We find that the conductivity increases with the frequency and temperature, as shown in Fig.~\ref{ACS} for all samples. The dispersive type behavior at lower frequencies is due to the accumulation of space charge inside the material, which vanishes at higher frequencies and temperatures. The conductivity starts to increase (due to enhanced hopping of carriers) abruptly above hoping frequency due to the short-range hopping of polarons, as the capacitor impedance becomes lower than the resistor impedance inside the material. Also, the conductivity increases with temperature for all samples due to the increased hopping probability of polarons due to an increase in activation energy and a reduction in barrier heights. The frequency dependence of {\it a.c.} conductivity data are explained using the jump relaxation model; according to this model, at lower frequencies, the charge carrier jumps from one site to another vacant site, giving the long-range translational motion of carriers, providing the frequency-independent conductivity contributing as {\it d.c.} conductivity. The shorter period available at higher frequency gives the two competing processes, unsuccessful and successful hopping and the ratio of these two processes creates the dispersive behavior of the {\it a.c.} conductivity \cite{Funke_PSSC_93, Sumi_Jap_10}. The frequency and temperature dependence of a.c. conductivity can be explained using the universal Jonscher power law: \cite{Raut_JAP_18, Jonscher_Nature_77, Sumi_Jap_10} 
\begin{equation} 
\label{power-law}
	\sigma_{(ac)} (\omega, T) = \sigma_{dc} (T) +A ~\omega^s
	\end{equation}
here, the $\sigma_{dc} (T)$ is the frequency-independent contribution in total conductivity called {\it d.c.} conductivity is obtained by extrapolating the low-frequency region to zero frequency attributed to the long-range translational motion of charge carriers, The term A$\omega^s$ gives the dispersion behavior in the conductivity, whereas $A$ is the temperature-dependent coefficient that measures the strength of polarization arises due to diffusive motion of charge carriers and $s$ is a parameter that characterizes the interaction between the carriers and its surrounding medium. The values of the $s$ parameter vary between 0 to 1 where the $s=$1 represents no interaction between the lattice and charge carriers, while a decrease of $s$ from 1 indicate the significant interaction between the lattice and mobile ions \cite{Nasari_CI_16, Nasari_CI_16, Jonscher_Nature_77, Sumi_Jap_10}. The temperature dependent variations of exponent $s$ provide information about the types of conduction; for example, if the $s$ value increases or decreases with the temperature, the conduction is governed by the small polaron hopping or correlated barrier hopping (CBH) \cite{Ghosh1_PRB_90, Taher_MRB_16, Mollh_JAP_93}. If the $s$ parameter decreases with temperature and attains a minimum, and then it increases, the conduction is governed by the overlapping large polaron tunneling (OLPT) or if the $s$ parameter attains a value nearly equal to 0.8, the conduction is governed by the quantum mechanical tunneling (QMT) \cite{Long_AIP_82, Kotkata_JAPD_06, Ghosh_PRB_90}. We have fitted the isothermal frequency dependence of {\it a.c.} conductivity using the modified power law: \cite{Bechir_JAP_14, Karmakar_JPCM_19, Megdiche_JALCOM_14}
\begin{equation} 
\label{MPL}
\sigma_{(ac)} (\omega) = \frac{\sigma_l}{(1+\omega^2 \tau^2)}+\frac{\sigma_{\infty} \tau^2 \omega^2 }{1+\omega^2 \tau^2} +A ~\omega^s
\end{equation}
where $\sigma_l$ and $\sigma_{\infty}$ represent the conductivity at low and high-frequency limits, $\tau$ is the characteristics relaxation time, $A$ and $s$ have the same usual meaning as having in the power law. The fitted results using the equation \ref{MPL} are shown in Fig.~\ref{ACS} that show that the $s$ parameter increases, see Fig.~5 of \cite{SI} with an increase in temperature, suggesting the correlated barrier hopping type conduction for all the samples in the measured temperature range. In the correlated barrier hopping (CBH) model, the conduction carriers occur between two defect sites by (a) single polaron hopping and (b) bipolaron hopping. 

\section{\noindent ~Conclusions}

 In summary we have successfully synthesized polycrystalline Na$_{3+x}$Zr$_{2-x}$Pr$_{x}$Si$_2$PO$_{\rm 12}$ ($x=$ 0.05--0.2) using the solid-state reaction and investigate their structural, morphological, dielectric relaxation, impedance behavior, electric modulus, and {\it a.c.} conductivity. The Rietveld refinement confirms the monoclinic (space group C2/c) phase with a small amount of ZrO$_2$ and Na$_3$Pr(PO$_4$)$_3$ impurity phase for high doping samples. The SEM analysis shows the distribution of the particles throughout the volume having various grain and grain boundaries as micro-constituents with a reduced porosity with Pr doping. The elemental composition analysis using EDX confirms the absence of unwanted elements. The temperature-dependent variation in dielectric constant and dielectric loss is explained using the thermal activation of charge carriers. The frequency dependence of electric permittivity and loss are explained using MWS polarization with space charge/Interfacial polarization model.The real and imaginary parts of electric permittivity are fitted using a modified cole-cole equation, confirming the non-Debye type relaxation in the measured temperature range. The broad impedance variation with frequency at selected temperatures show grain boundary and grain related relaxation at lower and high frequencies, respectively. The real and imaginary modulus fitting shows the distribution of relaxation times having non-Debye type relaxation as proposed by KWW function. The scaling and relaxation peak analysis shows similar types of relaxation for all Pr-doped samples. The {\it a.c.} conductivity data are fitted using modified power law, and the temperature dependence of the $s$ parameter shows the correlated barrier hopping (CBH) conduction in the measured temperature range. The higher values of electric permittivity at low frequencies make Pr-doped NASICON a suitable material for charge storage applications. 

\section{\noindent ~Acknowledgments}

RM thanks IUAC for providing the experimental facilities and Department of Physics of IIT Delhi for providing XRD facility. RSD acknowledges the financial support from SERB-DST through a core research grant (project reference no. CRG/2020/003436).

\end{document}